\documentclass[pra,aps,twocolumn,showpacs]{revtex4}
\usepackage{times,epsfig,amssymb,amsfonts,amsmath}
\newcommand{\ket}[1]{|#1\rangle}
\newcommand{\bra}[1]{\langle #1|}
\newcommand{\proj}[1]{\ket{#1}\bra{#1}}

\begin{document}

\title{Entanglement evolution of three-qubit mixed states in multipartite
cavity-reservoir systems}

\author{Jing-Zhou Xu $^{1}$}
\author{Jin-Bao Guo $^1$}
\author{Wei Wen $^2$}
\author{Yan-Kui Bai $^1$}
\email{ykbai@semi.ac.cn}
\author{Fengli Yan $^1$}
\email{flyan@hebtu.edu.cn}

\affiliation{$^1$ College of Physical Science and Information
Engineering and Hebei Advance Thin Films Laboratory, Hebei Normal
University, Shijiazhuang, Hebei 050016, China\\
$^2$ State Key Laboratory for Superlattices and Microstructures,
Institute of Semiconductors, Chinese Academy of Sciences,
P. O. Box 912, Beijing 100083, China}

\begin{abstract}
We analyze the multipartite entanglement evolution of three-qubit mixed
states composed of a GHZ state and a W state. For a composite system
consisting of three cavities interacting
with independent reservoirs, it is shown that the entanglement evolution is
restricted by a set of monogamy relations. Furthermore, as quantified by the
negativity, the entanglement dynamical property of the mixed entangled state
of cavity photons is investigated. It is found that the three cavity photons
can exhibit the phenomenon of entanglement sudden death (ESD). However, compared
with the evolution of a generalized three-qubit GHZ state which has the equal
initial entanglement, the ESD time of mixed states is later than that of the
pure state. Finally, we discuss the entanglement distribution in the multipartite
system, and point out the intrinsic relation between the ESD of cavity photons
and the entanglement sudden birth of reservoirs.
\end{abstract}


\maketitle

\section{Introduction}

Quantum entanglement is one of the important nonclassical features of
quantum mechanics. In recent decades, it has been rediscovered as a crucial
resource in quantum information processing, such as quantum communication
and quantum computation \cite{nielsen00book}. Therefore, it is a fundamental
problem to characterize the entanglement property of quantum systems. Till now, although
the entanglement in bipartite systems is well understood in many aspects, the
corresponding property in multipartite systems is far from clear, even
for three-qubit quantum states \cite{horodecki09rmp}.

In three-qubit pure states, the two-qubit and genuine three-qubit entanglement
can be quantified by the concurrences \cite{wootters98prl} and the three-tangle
\cite{ckw00pra}, respectively, which constitute a good hierarchy structure
for the entanglement characterization. However, for three-qubit mixed states,
the case is much more complicated and different. Lohmayer \emph{et al} analyzed
the entanglement property of a three-qubit mixed state composed of a GHZ state
and a W state \cite{lohmayer06prl}. They pointed out that, in some region of the
mix probability, the mixed state is entangled but there is no concurrence and three-tangle.
Compared with the entanglement in pure states, this is a new type of entangled state which
exhibits in the qubit-block form. Furthermore, based on a purification scenario, it is
elaborated that the entanglement actually comes from the genuine multipartite
entanglement between the system and its environment \cite{byw08pra}. However, for a deep
understanding of the three-qubit entanglement, it is desirable to investigate further
the entanglement property in the mixed states and compare it with that of pure states.

Entanglement dynamical property is very important in practical quantum
information processing. This is because that entanglement always decays due to
the unwanted interaction between the system and its environment. Theoretical
studies have revealed that entanglement does not always decay in an asymptotic
way and it can disappear in a finite time, which is referred to as entanglement
sudden death (ESD) \cite{horodecki01pra,rajagopal01pra,sch03jmo,tingyu06prl}
and has been detected in photon and atom systems experimentally \cite{almeida07sci,
laurat07prl} (see also a review paper \cite{tingyu09sci} and the references therein).

Recently, Lopez \emph{et al} analyzed the entanglement dynamics of two cavities
interacting with independent reservoirs \cite{lopez08prl}. They showed that when
the cavity entanglement experiences the ESD phenomenon, the reservoir entanglement
suddenly and necessarily appears. Moreover, it is shown that the entanglement
evolution is restricted by a monogamy relation and the genuine multipartite
entanglement is involved in the dynamical procedure \cite{byw09pra,ykbai11ejpd}.
However, these papers only addressed the evolution with bipartite initial states, and
the case of multipartite initial states is awaited for further studies. In three-qubit
mixed states, the qubit-block entanglement \cite{lohmayer06prl} is a new kind of
entanglement type and its dynamical property is yet to be considered.

In this paper, we analyze the entanglement dynamical property of a class of three-qubit
mixed states composed of a GHZ state and a W state. For a composite system consisting
of three cavities with independent reservoirs, we first show that the entanglement
evolution obeys a set of monogamy relations. Then, we investigate the entanglement evolution
of three cavity photons, which can exhibit the ESD property. This property is further
compared with that of a generalized GHZ state. Finally, we discuss the entanglement
distribution in the multipartite system and give the relation of entanglement between
the cavity photons and the reservoirs.

\section{Entanglement monogamy relations in multipartite cavity-reservoir systems}

Before deriving the monogamy relation in the dynamical evolution, we first introduce
the cavity-reservoir system. The interaction between
a single cavity and its N-mode reservoir is characterized by the Hamiltonian
\cite{lopez08prl}
\begin{equation}\label{1}
\hat{H}=\hbar \omega
\hat{a}^{\dagger}\hat{a}+\hbar\sum_{k=1}^{N}\omega_{k}
\hat{b}_k^{\dagger}\hat{b}_k+\hbar\sum_{k=1}^{N}g_{k}(\hat{a}
\hat{b}_{k}^{\dagger}+\hat{b}_{k}\hat{a}^{\dagger}).
\end{equation}
When a cavity mode contains a single photon and its
corresponding reservoir is in the vacuum state, the quantum state
is $\ket{\phi_0}=\ket{1}_{c}\otimes \ket{0}_{r}$, in which
$\ket{0}_{r}=\prod_{k=1}^{N}\ket{0_k}_{r}$. In the limit of $N\rightarrow \infty$,
the dynamical evolution given by the Hamiltonian in Eq. (1) leads to the output state
\begin{equation}\label{2}
\ket{\phi_t}=\xi(t)\ket{1}_c\ket{0}_r+\chi(t)\ket{0}_c\ket{1}_r,
\end{equation}
where $\ket{1}_r=(1/\chi(t))\sum_{k=1}^{N}\lambda_k(t)\ket{1_k}_r$ with the amplitude
converging to $\chi(t)=[1-\mbox{exp}(-\kappa t)]^{1/2}$ \cite{lopez08prl}. Thus,
the cavity and reservoir evolve as an effective two-qubit system.

We consider three cavities being affected by the dissipation of
three independent reservoirs. The initial state of three cavity photons and the
corresponding reservoirs is
\begin{equation}\label{3}
    \rho_{c_1c_2c_3r_1r_2r_3}(0)=\rho_{c_1c_2c_3}\otimes \proj{000}_{r_1r_2r_3},
\end{equation}
where the reservoirs are in the vacuum state and the cavity photons are in
the new type of entangled state \cite{lohmayer06prl}
\begin{equation}\label{4}
    \rho_{c_1c_2c_3}(0)=p\proj{GHZ}+(1-p)\proj{W},
\end{equation}
where $\ket{GHZ}=(\ket{000}+\ket{111})/\sqrt{2}$ and $\ket{W}=(\ket{001}+
\ket{010}+\ket{100})/\sqrt{3}$, respectively. In order to conveniently obtain the
output state of the multipartite cavity-reservoir systems, we introduce an ancillary
qubit $z$ which can purify the input state in Eq. (3). So, the global initial state
is
\begin{eqnarray}\label{5}
    \ket{\Psi_0}&=&(\sqrt{p}\ket{GHZ}\ket{0}+\sqrt{1-p}\ket{W}
    \ket{1})_{c_1c_2c_3z} \nonumber\\
    && \times \ket{000}_{r_1r_2r_3}.
\end{eqnarray}
Under the time evolution, the global output state has the form
\begin{eqnarray}\label{6}
  \ket{\Psi_t}&=&\sqrt{p/2}(\ket{000000}+\ket{\phi_t}
  \ket{\phi_t}\ket{\phi_t})_{c_1r_1c_2r_2c_3r_3}\ket{0}_{z}\nonumber\\
 && +\sqrt{(1-p)/3}(\ket{0000}\ket{\phi_t}+\ket{00}\ket{\phi_t}\ket{00}\nonumber\\
 && +\ket{\phi_t}\ket{0000})_{c_1r_1c_2r_2c_3r_3}\ket{1}_{z},
\end{eqnarray}
where $\ket{\phi_t}=\xi(t)\ket{10}+\chi(t)\ket{01}$ with $\xi(t)=\mbox{exp}(-\kappa t/2)$
and $\chi(t)=(1-\xi(t)^2)^{1/2}$. For the three cavity-reservoir system, its output state
is
\begin{equation}\label{7}
    \rho(t)_{c_1r_1c_2r_2c_3r_3}=\mbox{Tr}_{z}(\proj{\Psi_t}).
\end{equation}

In many-body quantum systems, entanglement is monogamous, which means that
entanglement cannot be freely shared among many parties. For a three-qubit state
$\rho_{ABC}$, Coffman \emph{et al} prove the relation \cite{ckw00pra}
\begin{equation}\label{8}
    C_{A|BC}^2\geq C_{AB}^2+C_{AC}^2,
\end{equation}
where the concurrence $C_{A|BC}^2=4\mbox{min}[\mbox{det}(\rho_A)]$
quantifies the bipartite entanglement in the partition $A|BC$ (the minimum runs over
all the pure state decompositions of $\rho_{ABC}$) and the concurrence $C_{ij}^2$
quantifies the two-qubit entanglement which is defined as
$C(\rho_{ij})=\mbox{max}(0,
\sqrt{\lambda_1}-\sqrt{\lambda_2}-\sqrt{\lambda_3}-\sqrt{\lambda_4})$
with the decreasing nonnegative real numbers $\lambda_{i}$ being the
eigenvalues of the matrix
$R_{ij}=\rho_{ij}(\sigma_y\otimes\sigma_y)\rho_{ij}^{\ast}
(\sigma_y\otimes\sigma_y)$ \cite{wootters98prl}. The monogamy relation in Eq. (8) is further
generalized to the $N$-qubit quantum state, which has the form
$C_{A_1|A_2\cdots A_N}^2 \geq C_{A_1A_2}^2+\cdots +C_{A_1A_N}^2$ \cite{osborne06prl}.
Moreover, for the two-qubit partition $A_1A_{1^\prime}|A_2A_{2^\prime}\cdots A_{N}A_{N^\prime}$,
a set of monogamy relations are also proved in the Ref. \cite{byw09pra}. But, these
relations are mainly focused on the distribution of two-qubit entanglement, the monogamy relation for
multiqubit entanglement still awaits further study.

In the dissipative procedure of the multipartite cavity-reservoir systems, we can
derive the following relations
\begin{subequations}
\begin{eqnarray}\label{9}
  C_{c_1|c_2c_3z}^{2}(0) &=& C_{c_1r_1|c_2r_2c_3r_3z}^2(t)\\
  &\geq& C_{c_1|c_2r_2c_3r_3z}^2(t)+C_{r_1|c_2r_2c_3r_3z}^2(t)\\
  &\geq& C_{c_1|c_2c_3}^2(t)+C_{r_1|r_2r_3}^2(t)\\
  &\geq& N_{c_1|c_2c_3}^2(t)+N_{r_1|r_2r_3}^2(t),
\end{eqnarray}
\end{subequations}
where the concurrence $C^2$ and the negativity $N^2$ \cite{vidal02pra} quantify the
bipartite entanglement in different multi-qubit quantum systems. In Eq. (9a),
we use the property that entanglement is invariant under local unitary (LU)
operations and consider the tensor product structure of the dynamical evolution
$U(\hat{H},t)=U_{c_1r_1}(\hat{H},t)\otimes U_{c_2r_2}(\hat{H},t)\otimes U_{c_3r_3}(\hat{H},t)$.
In  Eq. (9b), we use the monogamy relation of three-qubit quantum states, where
the subsystem $c_2r_2c_3r_3z$ can be regarded as a logic qubit. In  Eq. (9c),
we use the property that entanglement does not increase under local operations and classical
communication (LOCC), and we trace out the subsystems $r_2r_3z$ and $c_2c_3z$, respectively.
In  Eq. (9d), we use the relation $C(\rho_{AB})\geq N(\rho_{AB})$ \cite{kchen05prl,hfan07pra},
in which the negativity is defined as \cite{vidal02pra}
\begin{equation}\label{10}
    N(\rho_{AB})=||\rho^{T_A}||-1,
\end{equation}
where $\rho^{T_A}$ is the partial transpose with respect to the subsystem $A$, and the norm is
$||R||=\mbox{Tr}\sqrt{R^\dagger R}$.

According to Eq. (9), we know that the initial entanglement
between the cavity photon $c_1$ and the other photons $c_2c_3$ plus the ancillary system $z$
restricts the entanglement evolution in the multipartite cavity-reservoir systems. This
entanglement is not less than the sum of concurrences $C_{c_1|c_2r_2c_3r_3z}^2$ and
$C_{r_1|c_2r_2c_3r_3z}^2$. Furthermore, this equation gives hierarchy relations of entanglement
in three cavity photons and three reservoirs. About the set of monogamy relations,
we want to point out two points. The first one is that the monogamy relations are not limited by the
initial state in Eq. (5) and they are satisfied only if the initial cavity photons are in
a three-qubit quantum state and the reservoirs in the vacuum state. The second one is that
the relation $C_{c_1|c_2c_3}^2(0)=C_{c_1r_1|c_2r_2c_3r_3}^2(t)$ is satisfied in the
dynamical procedure, however, the similar monogamy relations in Eq. (9) do not hold in general.

The concurrence for multiqubit mixed states is
difficult to compute, because its solution requires to run over all the pure state decomposition
and take the minimum. Compared with the concurrence, the negativity is computable and we
will use this measure to analyze the entanglement evolution in the multipartite
cavity-reservoir systems.

\section{Entanglement evolution of three-qubit mixed states}

In the multipartite cavity-reservoir system, the initial state of the three
cavity photons is the mixed state composed of a GHZ state and a W state as shown in
Eq. (4). This quantum state has a special entanglement property that, in a certain region of
the mix probability, it is entangled but the entanglement can not be explained as the two-
qubit concurrences or the three-qubit tangle \cite{ckw00pra,lohmayer06prl,byw08pra}.
It has a new type of entanglement structure and can exhibit in the qubit-block form
\cite{byw09pra}. In the following we will first analyze its entanglement evolution property
in the multipartite cavity-reservoir system, and then compare it with
a generalized GHZ state.

In the dissipative environment, the output state of the three cavity photons
can be written as
\begin{equation}\label{11}
    \rho(t)_{c_1c_2c_3}=\mbox{Tr}_{r_1r_2r_3z}(\proj{\Psi_t}),
\end{equation}
where $\ket{\Psi_t}$ is the global evolution state given in Eq. (6). We use negativity
to characterize the entanglement evolution in the dynamical procedure. For the output
state of cavity photons, its negativity is
\begin{equation}\label{12}
    N[\rho(t)_{c_1c_2c_3}]=||\rho^{T_{c_1}}(t)||-1=\sum_i |\lambda_i(t)|-1,
\end{equation}
where $\lambda_i$s are the eigenvalues of the matrix $\rho^{T_{c_1}}$.
In the time evolution, the entanglement degradation of cavity photons has two kinds of routes,
one is the asymptotic way, the other is the ESD way. The entanglement is
nonzero when at least one eigenvalues is negative, and it is zero when all
the eigenvalues are nonnegative. After some calculation, we can obtain the eigenvalues of the
matrix $\rho^{T_{c_1}}(t)$, which can be expressed as
\begin{eqnarray}\label{13}
\lambda_{1} &=&\frac{1}{2}e^{-3\kappa t}p, \nonumber\\
\lambda_{2} &=&\frac{1}{2}e^{-3\kappa t}(e^{\kappa t}-1)p,\nonumber\\
\lambda_{3} &=&\frac{1}{2}e^{-3\kappa t}(e^{\kappa t}-1)^{2}p,\nonumber\\
\lambda_{4} &=&\frac{1}{6}e^{-\kappa t}[4-4p+3(1-e^{-\kappa t})^2p]\nonumber\\
\lambda_{5} &=&\frac{1}{12}[A_{1}-e^{-3\kappa t}\sqrt{A_{2}}],\nonumber\\
\lambda_{6} &=&\frac{1}{12}[A_{1}+e^{-3\kappa t}\sqrt{A_{2}}],\nonumber\\
\lambda_{7} &=&\frac{1}{12}[B_{1}-e^{-2\kappa t}\sqrt{B_{2}}],\nonumber\\
\lambda_{8} &=&\frac{1}{12}[B_{1}+e^{-2\kappa t}\sqrt{B_{2}}],
\end{eqnarray}
where the parameter $p$ is the mix probability of the GHZ and W states, $\kappa t$
represents the time evolution ($\kappa$ is the dissipative constant), and the
parameters $A_1$, $A_2$, $B_1$ and $B_2$ have the forms
\begin{eqnarray*}
A_{1}&=&e^{-\kappa t}(2-2p+3(1-e^{-\kappa t})p),\nonumber\\
A_{2}&=&18e^{3\kappa t}p(p-2)+36p^{2}-108e^{\kappa t}p^{2}\nonumber\\
      && +e^{4\kappa t}(p+2)^{2}+3e^{2\kappa t}p(8+31p),\nonumber\\
B_{1}&=&3[p+(1-e^{-\kappa t})^3p+(1-e^{-\kappa t})\nonumber\\
        &&\times (2-2p+e^{-2\kappa t}p)],\nonumber\\
B_{2}&=&36[e^{4\kappa t}+p^{2}-e^{3\kappa t}(p+2)-e^{\kappa t}p(p+2)] \nonumber\\
      && +e^{2\kappa t}(68+44p+41p^{2})\nonumber.
\end{eqnarray*}
In the eight eigenvalues, $\lambda_1$, $\lambda_2$, $\lambda_3$, $\lambda_4$,
$\lambda_6$, and $\lambda_8$ are always nonnegative, because they can be written as the product
or sum of some nonnegative terms. The eigenvalues $\lambda_5$ and $\lambda_7$ can be negative,
and they are served as the indicators of the evolution routes. Set $\lambda_5=0$, we can
solve the relation between the parameters $p$ and $\kappa t$, which can be written as
\begin{equation}\label{14}
    p=\frac{2e^{2\kappa t}(-1+e^{\kappa t})}{3-9e^{\kappa t}+7e^{2\kappa t}+2e^{3\kappa t}}.
\end{equation}
For a given value of the parameter $p$, $\lambda_5(p)< 0$ when the time $\kappa t<\kappa t(p)$,
and $\lambda_5(p)>0$ when $\kappa t>\kappa t(p)$. Set $\lambda_7=0$, we can obtain
the relation
\begin{equation}\label{15}
    p=\frac{9-18e^{\kappa t}+17e^{2\kappa t}-3\sqrt{D}}
    {8e^{2\kappa t}-9e^{-2\kappa t}(1-4e^{\kappa t}+4e^{2\kappa t}-e^{3\kappa t})},
\end{equation}
where the parameter $D=17-68e^{\kappa t}+102e^{2\kappa t}-76e^{3\kappa t}
+25e^{4\kappa t}$. Similarly, given a value of $p$, $\lambda_7(p)<0$ when
$\kappa t< \kappa t(p)$, and $\lambda_7(p)>0$ when $\kappa t>\kappa t(p)$.

\begin{figure}
\begin{center}
\epsfig{figure=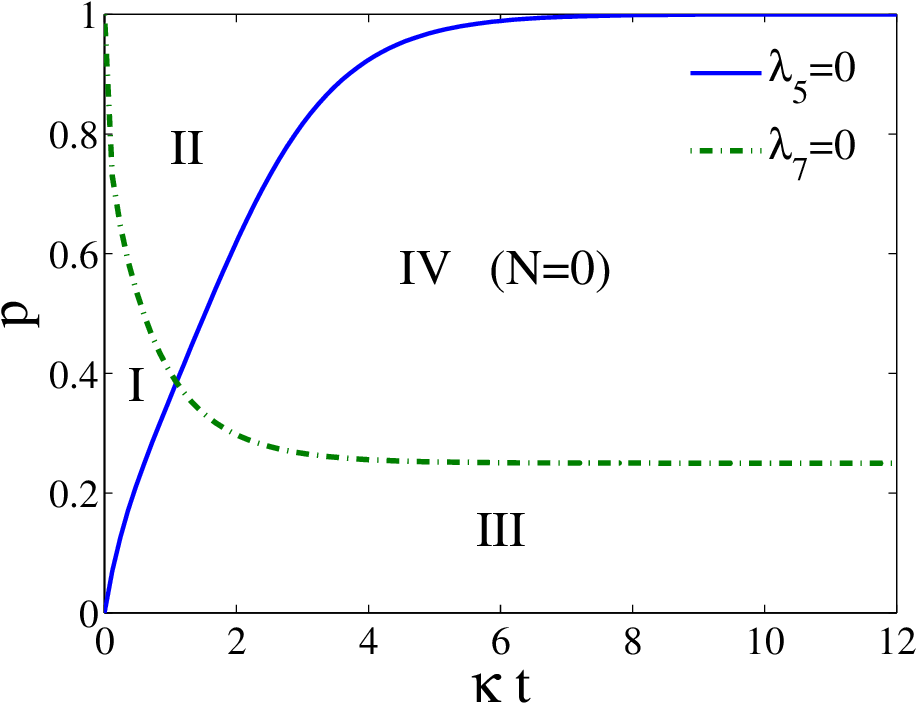,width=0.42\textwidth}
\end{center}
\caption{(Color online) Four entanglement evolution regions described
by the signs of eigenvalues $\lambda_5$ and $\lambda_7$. In the region IV,
both eigenvalues are positive, therefore the entanglement of three cavity
photons is zero in the dynamical procedure.}
\end{figure}

In Fig. 1, we plot the mix probability $p$ as a function of the time evolution $\kappa t$ when
$\lambda_5=0$  and $\lambda_7=0$, respectively. The two lines divide the whole area into
four parts where $\lambda_5<0$ and $\lambda_7<0$ in the region I, $\lambda_5<0$ and
$\lambda_7>0$ in region II, $\lambda_5>0$ and $\lambda_7<0$ in region III,
and, $\lambda_5>0$ and $\lambda_7>0$ in region IV, respectively.
According to the sign of the two eigenvalues, we know that the entanglement
of three cavity photons is zero in region IV and it is nonzero in the other three regions.
Therefore, the entanglement evolution experiences the ESD when the
corresponding mixed probability $p$ ranges in $(0.25,1)$, and the first ESD time occurs at
$\kappa t\simeq 1.091$ and the probability $p\simeq 0.385$. On the other hand, the entanglement
evolution is asymptotic when the probabilities $p\leq 0.25$ and $p=1$.

\begin{figure}
\begin{center}
\epsfig{figure=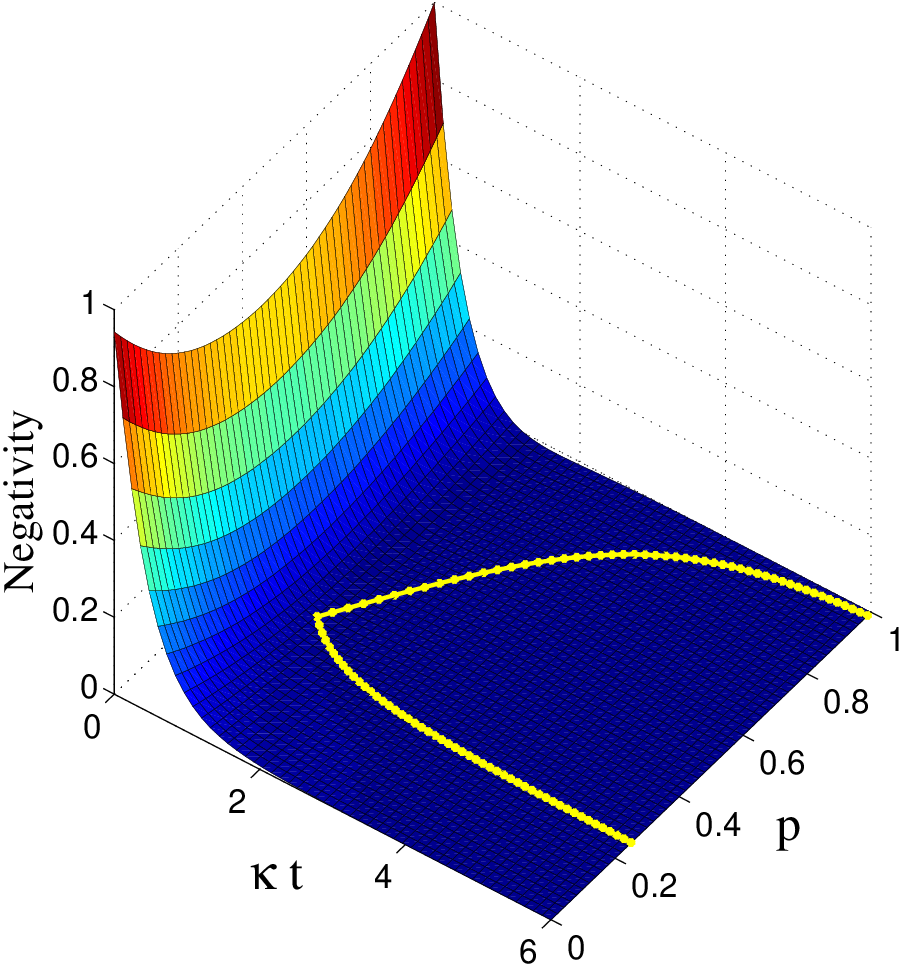,width=0.36\textwidth}
\end{center}
\caption{(Color online) Negativity of three cavity photons as a function of
the probability $p$ and the time evolution $\kappa t$, where the yellow line
characterizes the ESD time in the dissipative procedure.}
\end{figure}

In Fig.2, we plot the negativity of three cavity photons as a function of the mix probability
$p$ and the time evolution $\kappa t$. When $p=0$, the initial state is the W state
and its entanglement is $N=2\sqrt{2}/3$. Along with the time evolution, the entanglement decay
is asymptotic. With the increase of the parameter $p$, the initial state entanglement decreases
and the entanglement evolution is still asymptotic until the mix probability $p=0.25$. When
$p>0.25$, the entanglement decays in the ESD way except for $p=1$. The minimum of the
initial entanglement ($N\simeq 0.643$) is located at $p\simeq 0.465$,
however, its ESD time is not the least and the minimum of the ESD time corresponds to
the value $p\simeq 0.385$. When $p=1$, the initial state is the GHZ state ($N=1$) and its evolution
is asymptotic.

The three cavity photons are initially in the mixed state
composed of a GHZ state and a W state whose entanglement structure
is quite different from that of pure states. Lohmayer \emph{et al} pointed out that,
when the mix probability $p$ ranges in $[p_c, p_0]$ with
$p_c=7-\sqrt{45}\simeq 0.292$ and $p_0=4\sqrt[3]{2}/(3+4\sqrt[3]{2})\simeq 0.627$,
the mixed state is entangled but the entanglement is not the concurrence or the
three-tangle \cite{lohmayer06prl}. In this
range, the mixed state is still genuine three-qubit entanglement and exhibits in the qubit-block
form \cite{byw08pra}.
Therefore, it is desirable to analyze the difference of the entanglement dynamical property
between the mixed state and pure states. In the following, we will analyze the
entanglement evolution of a generalized GHZ state which is a genuine three-qubit entangled
state and has the form
\begin{equation}\label{16}
\ket{GHZ^g}=a\ket{000}+b\ket{111}.
\end{equation}
For the generalized GHZ state, its negativity $N(\ket{GHZ^g})=2ab$.
Its output state in the multipartite cavity-reservoir system has the form
\begin{equation}\label{17}
\rho_{c_1c_2c_3}^g(t)=\mbox{Tr}_{r_1r_2r_3}(\proj{\Psi^g(t)}),
\end{equation}
where $\ket{\Psi^g(t)}_{c_1r_1c_2r_2c_3r_3}=a\ket{000000}+b\ket{\phi_t}\ket{\phi_t}\ket{\phi_t}$
with $\ket{\phi_t}=\xi(t)\ket{10}+\chi(t)\ket{01}$ and $\xi(t)=\mbox{exp}(-\kappa t/2)$.
After some derivation, we can obtain that the negativity of the output state is
\begin{equation}\label{18}
N(\rho_{c_1|c_2c_3}^g)=\mbox{max}\{b e^{-3\kappa t}[\sqrt{F}-b e^{\kappa t}(e^{\kappa t}-1)],0\},
\end{equation}
where the parameter $F=4a^2\mbox{exp}(3\kappa t)+b^2[2-3\mbox{exp}(\kappa t)+\mbox{exp}(2\kappa t)]$.
Setting $N(\rho^g)=0$, we can further deduce the relation between the parameters $a$ and $\kappa t$,
which can be written as
\begin{equation}\label{19}
    a(\kappa t)=\sqrt{\frac{(e^{\kappa t}-1)^3}{(e^{\kappa t}-1)^3+e^{3\kappa t}}}.
\end{equation}
According to this relation, we can plot the ESD line in the entanglement evolution.
The generalized GHZ state can experience
the ESD phenomenon when the amplitude $a<\sqrt{2}/2$.
When $a\geq \sqrt{2}/2$, the entanglement evolution of quantum state $\rho^g_{c_1|c_2c_3}(t)$ is
asymptotic and the negativity is zero only for the limit of $\kappa t\rightarrow \infty$.

\begin{figure}
\begin{center}
\epsfig{figure=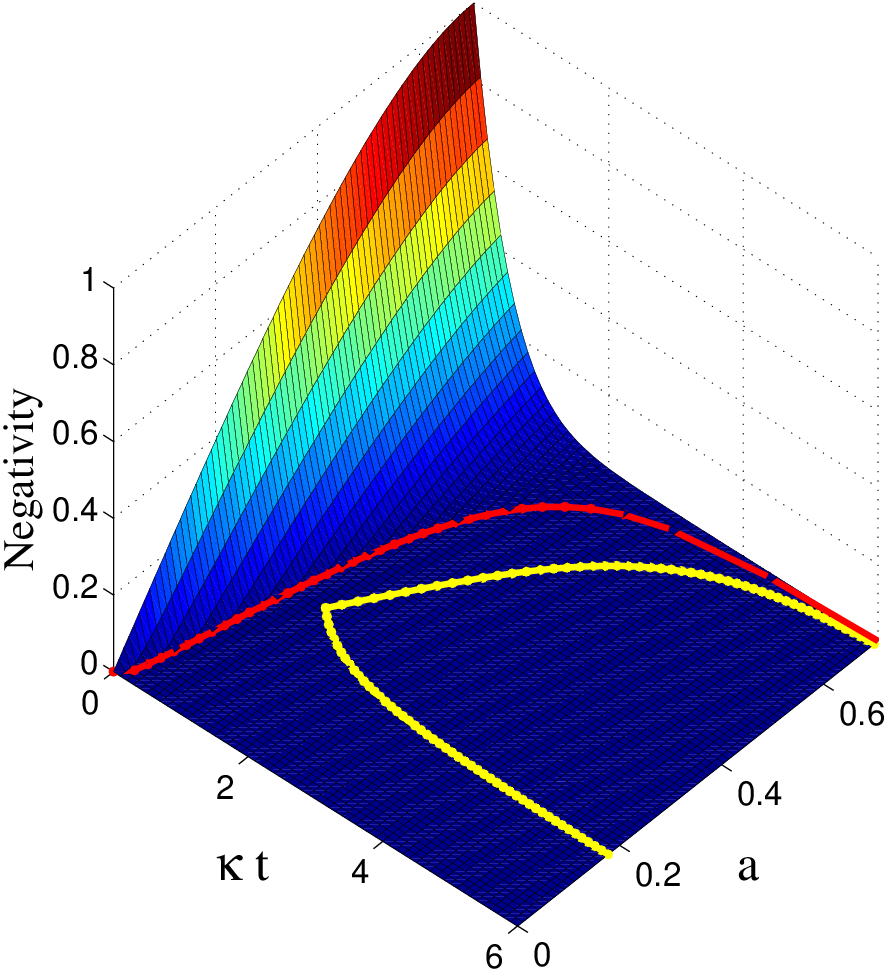,width=0.36\textwidth}
\end{center}
\caption{(Color online) Negativity of the generalized GHZ state, in which the
red line is its ESD line and the yellow line is the ESD line of the mixed state
composed of the GHZ and W states.}
\end{figure}

In Fig.3, we plot the negativity of the output state $\rho^g_{c_1c_2c_3}(t)$ as a function
of the amplitude $a$ and the time evolution $\kappa t$. The amplitude $a$ ranges in
$[0,\sqrt{2}/2]$ which corresponds to the negativity of the generalized GHZ state changing in $[0,1]$.
Along with the time evolution, the entanglement decays in the ESD way. The ESD line (the
red line) of the generalized GHZ state is plotted in the figure, where the ESD time increases
monotonically with the amplitude $a$. We also plot the ESD line (the yellow line) of
the mixed state composed of a GHZ state and a W state (the mix probability
$p$ is scaled with a constant $\sqrt{2}/2$ in order to be compatible with the amplitude $a$).
When the probability $p=1$, the mixed state becomes the GHZ state and it entanglement
evolution is asymptotic, which is the same as that of the generalized GHZ state for the case
$a=\sqrt{2}/2$. When $p\leq 0.25$, the evolution of the mixed state is still asymptotic,
however, in the same region of the amplitude $a$, the entanglement of the generalized GHZ state
decays in the ESD way. The mixed state has a special entanglement structure when the mix probability
$p\in [p_c,p_0]$. In the case that the generalized GHZ state has an equal initial entanglement to that of
the mixed state, we can derive that the parameter $a$ changes in $[0.319,0.363]$. In this region,
the maximal ESD time for the generalized GHZ state is $\kappa t\simeq 0.763$. However, the
minimal ESD time of the mixed state is $\kappa t\simeq 1.091$. Therefore, we have the conclusion,
with the equal initial entanglement, the ESD time of the mixed state is later than that of the
generalized GHZ state.

\section{discussion and conclusion}

A set of monogamy relations shown in Eq. (13) hold in the entanglement evolution,
which restrict the entanglement distribution. Particularly, in the dynamical procedure,
the initial entanglement between the cavity photon $c_1$ and the system $c_2c_3z$ is
larger than the sum of entanglement in subsystems $c_1c_2c_3$
and $r_1r_2r_3$. However, the monogamy relation given by
the initial entanglement $C_{c_1|c_2c_3}^2(0)$ has not been established.
The main reason is that, in this case, the related subsystems cannot
be regarded as a logic qubit. Ou pointed out that the monogamy relations induced by concurrences
do not hold in general for the higher dimensional objects \cite{ou07pra}.
However, entanglement monogamy is a fundamental property in many-body systems
\cite{kim09pra}, and the similar
monogamous relation can be satisfied under a well-defined entanglement measure.
With a good monogamy relation of entanglement, one can construct multipartite entanglement measures
\cite{ckw00pra,hfan07pra,byw07pra,baw08pra}, and have a deep understanding of many-body
quantum systems.

In the time evolution, the entanglement of three reservoirs have close relation to
that of cavity photons. After some calculation, we can obtain the relation of density matrices
between the two subsystems, which can be expressed as
\begin{equation}\label{20}
    \rho_{r_1r_2r_3}(\xi,\chi)=S_{\xi\leftrightarrow \chi}[\rho_{c_1c_2c_3}(\xi,\chi)],
\end{equation}
where $S$ is a transformation interchanging the parameters $\xi$ and $\chi$ with
$\xi=\mbox{exp}(-\kappa t/2)$ and $\chi=[1-\mbox{exp}(-\kappa t)]^{1/2}$. According to
this relation, we can know that the entanglement of three cavity photons will
transfer completely to the reservoir systems in the limit $\kappa t\rightarrow \infty$.
Moreover, when the ESD phenomenon of the entangled cavity photons occurs at
the time $t_{ESD}=t_0$,
the reservoir systems will experience the entanglement sudden birth (ESB) at the
time $t_{ESB}=-(1/\kappa t)\mbox{ln}[1-\mbox{exp}(-\kappa t_0)]$, which means that the ESD of
cavity photons and the ESB of reservoirs are intrinsically related in the multipartite system.

The entanglement evolution of three-qubit mixed states composed of a GHZ state
and a W state is analyzed, which has a better dynamical property than that of the generalized
GHZ state in the specific parameter regions. For a more general case,
consideration of the mix of
a generalized GHZ state shown in Eq. (16) and a generalized W state
$\ket{W^g}=\alpha\ket{100}+\beta\ket{010}+\gamma\ket{001}$ is deserved \cite{eosu08njp}.
The entanglement evolution has close relation to the type of enviroment, therefore the entanglement
dynamics in non-Markovian environments need to be considered in future \cite{bel99prl,jan07pra,
alt10pla}. Moreover,
beside the cavity-reservoir system, other physical systems are also worth being investigated,
for example, atom systems \cite{sch03jmo,tingyu06prl}, quantum dots \cite{loss98pra,ssli01pnas}
and spin chains \cite{waw06pra}.

In conclusion, we have analyzed the entanglement dynamical property of the mixed state composed
of a GHZ state and a W state in multipartite cavity-reservoir systems. As shown in Eq. (13),
the entanglement evolution is restricted by a set of monogamy relations. When the mix probabilities
$p\leq 0.25$ and $p=1$, the entanglement evolution of cavity photons is asymptotic, and,
for other region of the probability, the evolution is in the ESD way. Furthermore, in the
specific region $p\in[p_c,p_0]$, the ESD time of the mixed state is later than that of
the generalized GHZ state with the equal initial entanglement. Finally, we discuss the
entanglement distribution in the multipartite systems and point out the intrinsic relation
between the ESD of cavity photons and the ESB of reservoirs.

\section*{Acknowledgments}

This work was supported by the NSF-China under Grant No: 10905016 and the fund of
Hebei Normal University. F.L.Yan. was also supported by the NSF-China under Grant
No: 10971247, and Hebei Natural Science Foundation of China under Grant Nos: F2009000311,
A2010000344.

\end{document}